\documentclass{article}
\usepackage{microtype}
\usepackage{graphicx}
\usepackage{subfigure}
\usepackage{booktabs} 
\usepackage{enumerate}
\usepackage{hyperref}

\usepackage[accepted]{icml2025}

\hypersetup{colorlinks=true}
\usepackage{amsmath}
\usepackage{amssymb}
\usepackage{mathtools}
\usepackage{amsthm}

\DeclareMathOperator*{\argmin}{arg\,min}
\usepackage[capitalize,noabbrev]{cleveref}

\theoremstyle{plain}

\theoremstyle{definition}

\theoremstyle{remark}

\usepackage[textsize=tiny]{todonotes}

\begin{document}

\twocolumn[
    \icmltitle{Auto-WHATMD : Automated Wasserstein-based High-dimensional feature extraction Analysis of Trajectories from Molecular Dynamics}
    

    \begin{icmlauthorlist}
        \icmlauthor{Sosuke Asano}{keio}
        \icmlauthor{Ikki Yasuda}{keio}
        \icmlauthor{Katsuhiro Endo}{aist}
        \icmlauthor{Yoshinori Hirano}{keio}
        \icmlauthor{Kenji Yasuoka}{keio}
    \end{icmlauthorlist}

    \icmlaffiliation{keio}{Department of Mechanical Engineering, Keio University, Yokohama, Kanagawa, Japan}
    \icmlaffiliation{aist}{National Institute of Advanced Industrial Science and Technology, Tsukuba, Ibaraki, Japan}
    \icmlcorrespondingauthor{Kenji Yasuoka}{yasuoka@mech.keio.ac.jp}

    %\icmlkeywords{Protein, Molecular dynamics, Optimal transport, deep learning}
    \vskip 0.3in
]

\printAffiliationsAndNotice{}

\begin{abstract}
Comparing multiple protein systems with variation such as different binding ligands or mutations, and understanding their effects is one of the objectives in molecular dynamics simulations. 
Representation of these systems by a few features enables quantitative comparison. 
However, because molecular dynamics simulation trajectories are high-dimensional spatiotemporal data, selection of key features relies on domain expertise, sometimes introducing arbitrary assumptions.
Here, we present an approach that uses the optimal transport distance to compare high-dimensional trajectory data, and employs simulated annealing to identify the residues that best distinguish multiple systems.
We term this algorithm auto-WHATMD (automated \textbf{\underline{W}}asserstein-based \textbf{\underline{H}}igh-dimensional feature extraction \textbf{\underline{A}}nalysis for \textbf{\underline{T}}rajectories of \textbf{\underline{M}}olecular \textbf{\underline{D}}ynamics). 
We applied auto-WHATMD to multiple protein-ligand systems of bromodomain 4 with different ligands, identifying the most discriminative residues in the loop region. 
Moreover, even a few selected residues were sufficient to capture the correlation with ligand-binding affinities, indicating that auto-WHATMD effectively prioritizes the most informative residues.
Our approach can be used to efficiently determine key residues and design features for multiple analogous systems. 
\end{abstract}

\begin{figure*}[h!]
    %\vskip 0.2in
    \begin{center}
        \centerline{\includegraphics[width=160mm]{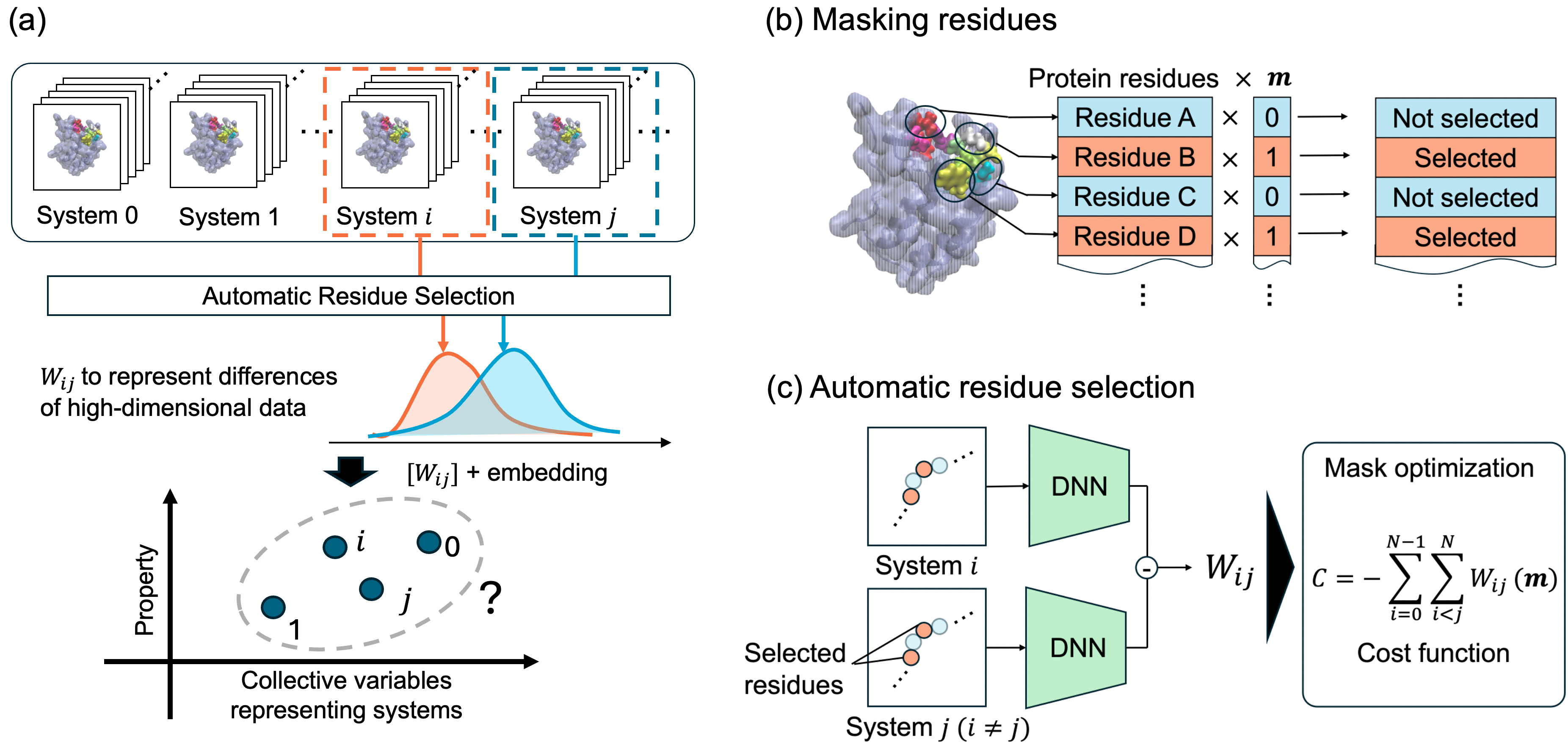}}
        \caption{
            Schematic overview of auto-WHATMD. 
            (a)~Multiple systems are prepared and important residues are automatically selected from the entire protein. 
            Trajectories of the selected residues are represented as high-dimensional data distributions. 
            Differences between systems are quantified using the Wasserstein distance, and the resulting distance matrix is embedded into a low-dimensional space to analyze relationships between collective variables and system properties.
            (b)~Important residues selected from the protein are indicated by residue masking.
            (c)~Optimization for automatic residue selection. System differences computed by a deep neural network are used to optimize the mask vector.
        }
        \label{fig:fig1_kai}
    \end{center}
    \vskip -0.2in
\end{figure*}

\section{Introduction}
Proteins have unique structures and dynamics with specific amino acid sequences. 
These structural and dynamic features are critical for their biological functions, including special tertiary structures that function as catalytic sites and ligand recognition sites or selective ion and ligand-dependent channels \cite{karplus2005molecular}, as well as flexible disordered regions that recognize other proteins and regulate self-assembly behaviors \cite{van2014classification}. 
However, many proteins have heterogeneous conformations, including metastable structures, in addition to the single structure accessible from X-ray crystal structure analysis data and prediction from machine learning models~\cite{bottaro2018biophysical}. This heterogeneity can range from minor scale, such as side-chain orientation change, to major-scale conformational changes, including open and close loop protein transitions \cite{henzler2007dynamic}.

Computational approaches for studying protein structural ensembles have been developed based on either physics-based methods, data-driven methods, or a combination of both. 
Molecular dynamics (MD) simulations, often starting from crystal structures, can capture structural changes and dynamics at atomistic or molecular resolution. 
They are widely employed to investigate protein dynamics and to sample heterogeneous conformational ensembles, providing valuable insights into protein folding, conformational changes, and ligand binding \cite{lindorff2011fast,plattner2015protein}.
Recently, machine learning approaches have been developed to predict static structures such as AlphaFold and similar methods \cite{jumper2021highly,baek2021accurate} and further being  extending to prediction of conformational ensembles~\cite{lewis2025scalable}. 
These methods include generative models such as flow-based models \cite{noe2019boltzmann}, generative adversarial networks \cite{janson2023direct}, and diffusion models \cite{lu2024dynamicbind,Lewis2024}.

Comparing structural ensembles is an important but challenging task~\cite{vogele2025systematic}. 
For evaluating a single structure, commonly employed approaches include contact maps, root mean square deviation (RMSD), t-SNE and relevant physical quantities~\cite{mardt2018vampnets, lemke2019encodermap,glielmo2021unsupervised}. 
Comparing ensembles is more challenging because they represent distributions of conformations rather than individual structures.
Further, molecular dynamics ensembles are generally time series of conformations, and therefore the ensemble is characterized by structural and temporal aspects, such as relaxation dynamics and diffusion~\cite{endo2018multi}. 
This motivates the need for systematic and objective methods to quantitatively compare conformational ensembles.

Optimal transport-based approaches have been developed and validated for comparing ensembles of biomolecules derived from MD simulations \cite{endo2019, yasuda2022, gonzalez2023wasco, mustali2023unsupervised}, alongside other comparison metrics (also see Related Work). 
However, feature selection plays a critical role in characterizing ensembles, as it significantly affects their representations. 
Trial and error is often required to achieve the desired representation; otherwise, inappropriate feature selection can lead to misleading interpretations.

In this work, we present a method for automatically selecting representative amino acid residues from the protein, and thereby comparing ensembles of multiple analogous systems. 
%Differences among the systems were obtained based on optimal transport, while the feature selection was stochastically processed using a simulated annealing approach. 
We applied this method to protein-ligand systems of a benchmarking protein bromodomain 4 (BRD4), using different ligand species bound to the protein. 
We demonstrated that our approach enables the successful identification of the crucial binding-site residues, and the differences in selected residues are associated with ligand-binding affinity.

\section{Related Work}
As a pioneering work in ensemble comparison, Brüschweiler extended the concept of RMSD to conformational ensembles and proposed the inter-ensemble RMSD (eRMSD) \cite{bruschweiler2003}. While this approach is both simple and useful, it has several limitations. For example, the eRMSD within the same ensemble is not zero and higher-order moments beyond the second moment are not considered. Consequently, it cannot fully capture the complexity of ensemble distributions.
Lindorff-Larsen et al. assumed that MD ensembles are sampled from the ground-truth distributions approximated as the sum of Gaussians \cite{lindorff2009similarity,tiberti2015encore}. 
They estimated these ensemble distributions and then computed metrics for comparison such as the Kullback-Leibler or Jensen-Shannon divergence. 
By applying dimensionality reduction to these matrices and projection onto 2D space, they further compared structural ensembles obtained from different experimental methodologies and MD simulation force fields \cite{lindorff2009similarity}. 
González-Delgado et al. utilized the Wasserstein distance \cite{gonzalez2023wasco}, which satisfies the axioms of a distance matrix and does not require any assumptions about the ground-truth distribution. They computed the Wasserstein distance for one-dimensional distributions of properties, such as the residue--residue distances and radius of gyration of disordered proteins, and demonstrated that it can differentiate conformational ensembles. 

Yasuda et al. proposed that the conformation dynamics of proteins are represented by an ensemble of short-term trajectories sampled from long MD trajectories, and that differences between two such ensembles can be quantified using the Wasserstein distance \cite{endo2019,yasuda2022}. 
Since the Wasserstein distance for high-dimensional distributions cannot be computed using linear algorithms due to its high computational cost \cite{villani2009optimal}, they approximated it using neural networks \cite{arjovsky2017,endo2019}. They demonstrated that multiple protein-ligand systems, differing only in ligand species, are projected into a feature space that correlates with binding affinity. However, this approach assumes that the systems are represented by a specific set of residues selected before computing the Wasserstein distance, which may introduce bias due to the potentially arbitrary choice of residues.
 
\section{Proposed Method}
Auto-WHATMD automatically performs feature extraction to differentiate multiple analogous systems, comprising three steps (Fig.~\ref{fig:fig1_kai}a): quantification of pairwise system differences based on the Wasserstein distance, automatic residue selection, and low-dimensional representation of the system differences. 
Previously, up to 20 residues were manually selected as input, ideally avoiding irrelevant residues~\cite{yasuda2022}. 
Auto-WHATMD automates this selection process, enabling automated analysis of multiple analogous systems.

We represent residue selection using a binary mask vector (Fig.~\ref{fig:fig1_kai}b), which indicates that the residue is included as input, and not included otherwise. 
Thus, the problem of residue selection is addressed by optimizing the mask vector. 
In the following sections, we present the methods to quantify the system differences, the optimization procedure of the mask, and a method to represent the pairwise system differences in low dimensions.

\subsection{Training neural network to calculate Wasserstein distance}~\label{sec:wd_nn}

Wasserstein distance is used as a measurement of distance between two distributions~\cite{villani2009optimal}, herein we use it to measure differences of two MD ensembles. 
For low-dimensional distributions, the Wasserstein distance can be calculated using linear programming with moderate computational cost. 
However, such methods are not feasible for high-dimensional distributions of short-time dynamics. 
Recently, neural networks have been used to approximate the Wasserstein distance in generative adversarial networks \cite{arjovsky2017}. 

We regard an MD system as a single distribution of high-dimensional data, referred to as the local dynamics ensemble, which consists of samples of short-term trajectories~\cite{endo2019}.
The local dynamics ensemble is useful for representing time-series data of many-particle systems in a data size suitable for neural network architectures.
One sample from the local dynamics ensemble has a dimensionality equal to the product of three factors (the time steps of the local dynamics, the number of selected residues, and the number of dimensions, which is three in $xyz$ space).

When the input is masked by a binary vector, the Wasserstein distance between the two local dynamics ensembles is defined as,
\begin{equation}
    \begin{aligned}
        W_{ij}(\boldsymbol{m}) =\sup_{\|f_{ij}\|_L \leq 1}\mathbb{E}_{\boldsymbol{x}_i \sim \boldsymbol{y}_i} [f_{ij}(\boldsymbol{m \odot x}_i)] \\
        - \mathbb{E}_{\boldsymbol{x}_j \sim \boldsymbol{y_j}} [f_{ij}(\boldsymbol{m \odot x}_j)]
        \label{eq:1-wd}
    \end{aligned}
\end{equation}
where $f_{ij}$ is a critic function represented by a deep neural network, and $\sup_{\|f_{ij}\|_L \leq 1}$ denotes the maximization over the space of functions satisfying the 1-Lipschitz constraint, $\mathbb{E}$ is the expectation,  and $\boldsymbol{x}_i$ is the data sampled from the distribution of local dynamics ensemble of system $i$, denoted by $\boldsymbol{y}_i$. The input data $\boldsymbol{x}_i$ is masked using the binary vector, $\boldsymbol{m}$, $\boldsymbol{m}_t$ and $\boldsymbol{m}_{\text{new}}$ are used to mask the input data. 
For simplicity, they are collectively represented as $\boldsymbol{m}$ in this explanation. 

The neural network function, $f_{ij}$, is trained for a specific pair of the local dynamics ensembles of interest. 
However, the mask is not yet determined at this stage of optimization of $f_{ij}$, and thus the local dynamics ensembles of residues to be selected are also not determined.
To address this, we randomly generate binary mask vectors for each step of training $f_{ij}$, under the assumption that $f_{ij}$ can output an approximation of the Wasserstein distance for varying mask vectors.

To satisfy the 1-Lipschitz constraint in Eq.~\ref{eq:1-wd}, $f_{ij}$ is trained using the WGAN-GP \cite{gulrajani2017improved} with the loss function $L$, 
\begin{equation}
\begin{aligned}
     L = \mathbb{E}_{\boldsymbol{x}_i \sim \boldsymbol{y}_i}[f_{ij}(\boldsymbol{m }\odot \boldsymbol{x}_i)] - \mathbb{E}_{\boldsymbol{x}_j \sim \boldsymbol{y}_j}[f_{ij} (\boldsymbol{m } \odot \boldsymbol{x}_j)] \\
     + \lambda \mathbb{E}_{\boldsymbol{r \sim R}}[\|\nabla_{\boldsymbol{m }\odot \boldsymbol{r}} f_{ij}(\boldsymbol{m }\odot \boldsymbol{r})\|-1]^2
\end{aligned}
\label{eq:loss}
\end{equation}
where $\boldsymbol{r}=\epsilon \boldsymbol{x}_i + (1-\epsilon)\boldsymbol{x}_j$, and $\epsilon$ is sampled from the uniform distribution in the range of $[0,1]$ for each step of training, $\boldsymbol{R}$ is probability distribution of $\boldsymbol{r}$. 
The third term, the gradient penalty, enforces the $1$-Lipschitz constraint on $f_{ij}$ by penalizing deviations of the gradient norm from unity along straight lines between samples drawn from $\boldsymbol{y}_i$ and $\boldsymbol{y}_j$. 
Specifically, the gradient is evaluated with respect to the masked input $\boldsymbol{m} \odot \boldsymbol{r}$, ensuring that the Lipschitz constraint is imposed in the masked feature space used for Wasserstein distance estimation. 
The coefficient $\lambda$ controls the strength of this regularization.

\begin{algorithm}[tb!]
\caption{Mask optimization}
\label{alg:auto-whatmd}
\begin{algorithmic}[1]
\INPUT Datasets $\{\boldsymbol{y}_i\}_{i=1}^N$, number of selected residues $n$, number of random search steps $N_{\text{rand}}$, maximum SA steps $N_{\text{SA}}$, $T_{\text{init}}$, $\gamma$, patience $K(=1000)$
\OUTPUT $\boldsymbol{m}^*$

\STATE \textbf{Phase 1: Initial random search}
\FOR{$b \gets 1$ \textbf{to} $N_{\text{rand}}$}
    \STATE Generate random mask $\boldsymbol{m}_b$ (with $n$ ones)
    \STATE Compute $C(\boldsymbol{m}_b)$ using Eq.~(\ref{eq:cost-func})
\ENDFOR
\STATE $\boldsymbol{m}_0 \gets \argmin_{b \in \{1,\dots,N_{\text{rand}}\}} C(\boldsymbol{m}_b)$

\STATE \textbf{Phase 2: Simulated annealing}
\STATE $t \gets 0$, $\boldsymbol{m}_t \gets \boldsymbol{m}_0$, $s \gets 0$
\FOR{$t \gets 1$ \textbf{to} $N_{\text{SA}}$}
    \STATE $T_t \gets \gamma^t T_{\text{init}}$
    \STATE Generate $\boldsymbol{m}_{\text{new}}$ by swapping adjacent $01 \leftrightarrow 10$
    \STATE Compute $C(\boldsymbol{m}_{\text{new}})$ using Eq.~(\ref{eq:cost-func})
    \STATE $\Delta C \gets C(\boldsymbol{m}_{\text{new}}) - C(\boldsymbol{m}_{t-1})$
    \IF{$\Delta C < 0$}
        \STATE $\boldsymbol{m}_t \gets \boldsymbol{m}_{\text{new}}$
    \ELSE
        \STATE $P_t \gets \exp\!\left(-\Delta C / T_t\right)$ \hfill (Eq.~(\ref{eq:metropolis_criteria}))
        \STATE $u \sim U(0,1)$
        \IF{$u < P_t$}
            \STATE $\boldsymbol{m}_t \gets \boldsymbol{m}_{\text{new}}$
        \ELSE
            \STATE $\boldsymbol{m}_t \gets \boldsymbol{m}_{t-1}$
        \ENDIF
    \ENDIF
    \IF{$C(\boldsymbol{m}_t) < \min_{k \in \{0,\dots,t-1\}} C(\boldsymbol{m}_k)$}
        \STATE $s \gets 0$
    \ELSE
        \STATE $s \gets s + 1$
    \ENDIF
    \IF{$s \ge K$}
        \STATE \textbf{break}
    \ENDIF
\ENDFOR

\STATE $\mathcal{M}_{\text{visited}} \gets \{\boldsymbol{m}_t\}_{t=1}^{N_{\text{SA}}}$
\STATE $\boldsymbol{m}^* \gets \argmin_{\boldsymbol{m} \in \mathcal{M}_{\text{visited}}} C(\boldsymbol{m})$
\end{algorithmic}
\end{algorithm}

\subsection{Mask vector optimization for residue selection} 
Using the neural networks trained to estimate the Wasserstein distance using the random mask, we optimize the binary mask vector. 
The algorithm for optimizing residue selection consists of two phases based on simulated annealing (Fig.~\ref{fig:fig1_kai}c and Algorithm~\ref{alg:auto-whatmd}).
%The Wasserstein distance measures the distance between two distributions of the local dynamics ensembles, defined as an ensemble of short-term time-series of the displacements of the selected residue's center of mass. 
%In Phase 1, we optimized the mask that specifies the selection of residues (Fig.~\ref{fig:fig1_kai}b using simulated annealing (Fig.~\ref{fig:fig1_kai}c). 
%To identify the subset of residues that best characterizes the differences between the systems, we optimize the residue-selection mask (Fig.~\ref{fig:fig1_kai}b) using a two-phase procedure based on simulated annealing (Fig.~\ref{fig:fig1_kai}c).

In Phase 1, we performed the initial random search, in which the promising parameter space of mask vectors is explored using the randomly generated mask vectors \cite{Li_2012}. Each mask vector is evaluated based on the mask optimization cost function $C$. Since larger Wasserstein distances indicate better discrimination between systems, we define the cost function as the negative sum of the pairwise Wasserstein distances, thereby formulating the optimization as a minimization problem. The cost function $C$ is defined as,
\begin{equation}
    C(\boldsymbol{m}) = -\sum_{i=1}^{N-1} \sum_{i < j}^N W_{ij}(\boldsymbol{m})
    \label{eq:cost-func}
\end{equation}
where $N$ is the number of systems, $\boldsymbol{m}$ is a binary mask vector, and $W_{ij}$ is the Wasserstein distance between system $i$ and $j$. We randomly generate mask vectors and then select one that minimizes $C$. The best-performing mask in Phase 1 is then used as the initial binary mask in Phase 2.

In Phase 2, we generate a new mask by swapping neighboring one and zero, that is, "01" is swapped to "10" or vice versa. We evaluated the generated mask using $C(\boldsymbol{m}_{\text{new}})$, accepting it according to the Metropolis criterion at probability $P_t$,
\begin{align}
\Delta C &= C(\boldsymbol{m}_{\text{new}}) - C(\boldsymbol{m}_{t-1}), \\
P_t &=
\begin{cases}
1, & \text{if } \Delta C < 0, \\
\exp\!\left(-\dfrac{\Delta C}{T_t}\right), & \text{if } \Delta C \ge 0.
\end{cases}
\label{eq:metropolis_criteria}
\end{align}
where $t$ denotes the current optimization step, $C(\boldsymbol{m}_{t-1})$ and $C(\boldsymbol{m}_{\text{new}})$ are the cost values for the current and new generated masks, respectively, and $T_t$ is the temperature at step $t$. The temperature is exponentially decayed as,
\begin{equation}
    T_t=\gamma^t T_{\text{init}}
\end{equation}
where $T_{\text{init}}$ is the initial temperature, $\gamma \in (0, 1)$ is the cooling rate. 

During simulated annealing, the value of the cost function $C$ is recorded at each step. 
Rather than selecting the mask obtained at the final iteration, we retain the mask that attains the minimum cost value over the entire optimization trajectory. 
Specifically, the final selected mask $\boldsymbol{m^{*}}$ is defined as
\begin{equation}
\boldsymbol{m}^* = \argmin_{\boldsymbol{m} \in \mathcal{M}_{\text{visited}}} C(\boldsymbol{m})
\end{equation}
Where $\mathcal{M}_{\text{visited}}$ denotes the set of masks explored during simulated annealing. This strategy ensures that the best-performing mask encountered during the search process is used for subsequent analysis.
% We used initial temperature 10.0 and cooling rate 0.999, and the optimization were performed for $N_3$ steps ($N_3=2,500$).  The total number of selected residues, $n$, is a hyperparameter and is constant during the mask optimization. 

%we ensure that the resulting Wasserstein distances from using random mask and fixed mask show a high correlation. 

\subsection{Feature extraction from the matrix of Wasserstein distance}
After computing the Wasserstein distances for all pairs of systems, we embedded this matrix of Wasserstein distances into the low-dimension space (Fig.~\ref{fig:fig1_kai}a). We used non-linear dimension reduction using simulated annealing and the following gradient descent method \cite{endo2019, yasuda2022}. The embedding points are adjusted so that their pairwise Euclidean distance approximates the Wasserstein distances, thus the embedded points are defined as,
\begin{equation}
    \{\boldsymbol{p}_i\}_{i=1}^N = \underset{\{\boldsymbol{p}_i\}_{i=1}^N}{\arg\min} \sum_{1\leq i<j\leq N} \left( W_{ij}(\boldsymbol{m^*}) - \|\boldsymbol{p}_i - \boldsymbol{p}_j\| \right)^2
    \label{eq:embed}
\end{equation}
Subsequently, principal component analysis was performed on the embedded points to ensure rotational invariance. 

%\vspace{20pt}
%\begin{table}[b]
%\centering
%\caption{Simulated annealing settings in this study}
%\begin{tabular}{cc}
%\hline
%\textbf{Settings}          & \textbf{}              \\ \hline
%Number of steps in method 1                  & 2500 \\
%Number of steps in method 2                  & 2500 \\
%Initial temperature              & 10.0        \\
%Cooling rate    & 0.999                     \\ \hline
%\end{tabular}
%\label{tab:tab1}
%\end{table}

\begin{figure}[tb]
    \includegraphics[width=\linewidth]{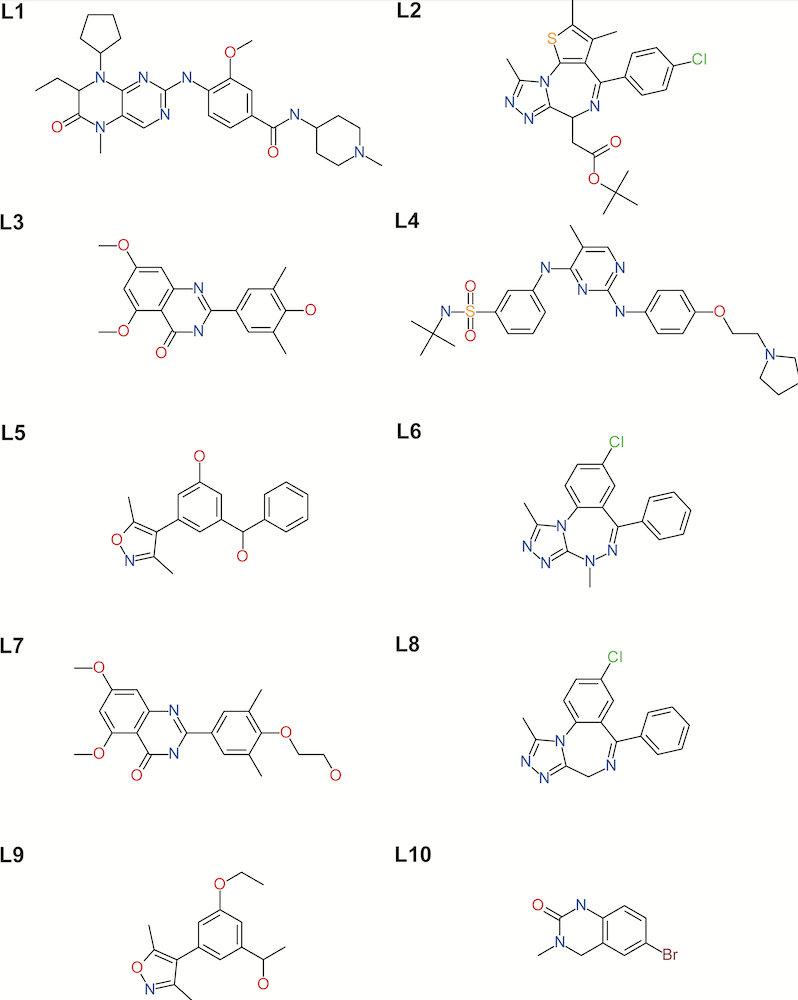}
    \caption{Chemical structure of ligand L1-10}
    \label{fig:ligand_image}
\end{figure}  
\begin{figure}[tb]
    \includegraphics[width=80 mm]{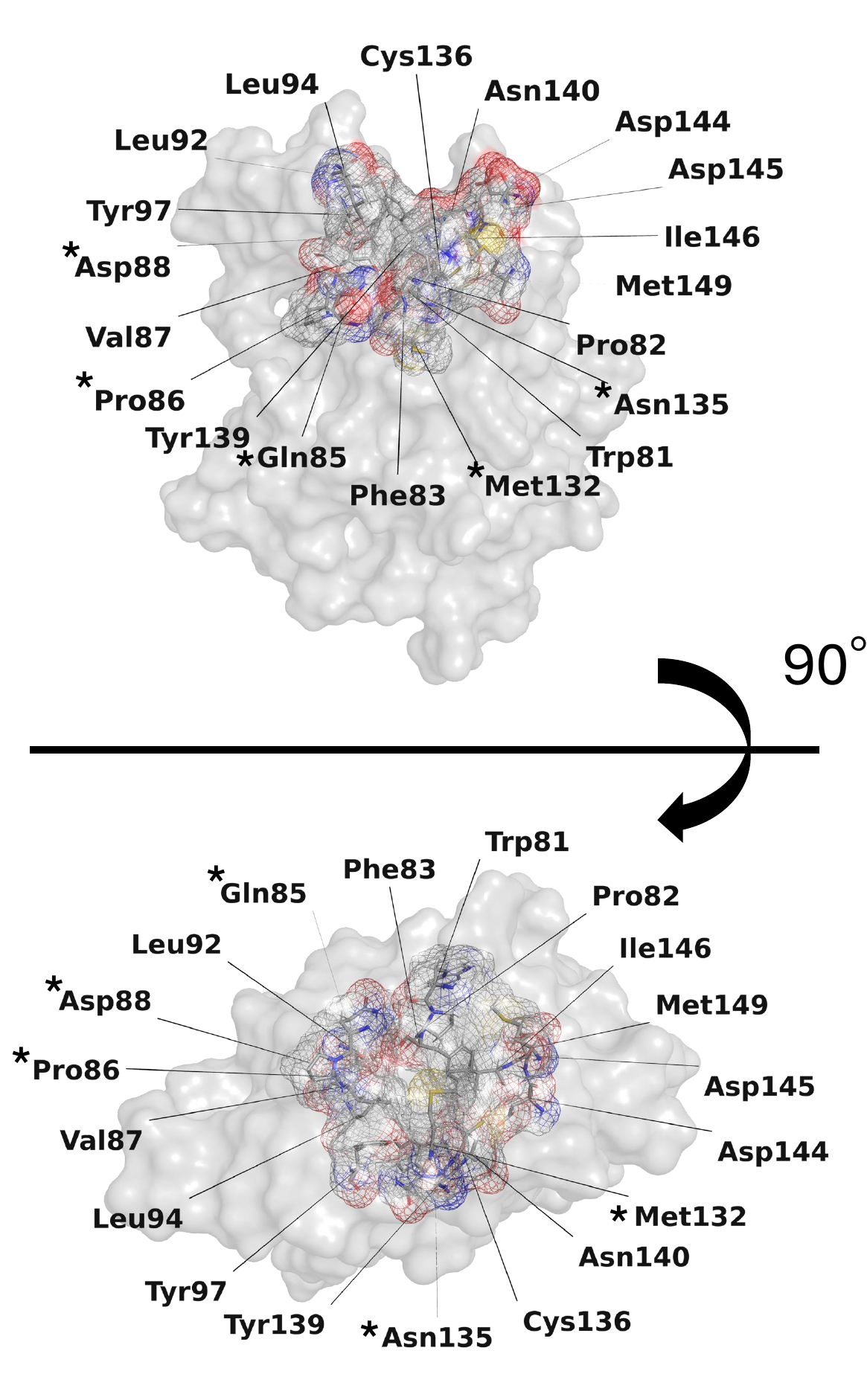}
    \caption{
    Spatial distribution of residues around the BRD4 binding site. The upper and lower panels show two views rotated by 90°. 
    Residues in the small subset (14 residues) are labeled, and those additionally included in the extended binding-site subset are marked with an asterisk (*).  
    }
    \label{fig:brd4_image}
\end{figure}  

\section{Experiment Settings}

\subsection{Dataset}
To assess the performance of Auto-WHATMD in identifying key residues from MD trajectories of biomolecules, we prepared a dataset of protein--ligand complex systems from a previous study \cite{yasuda2022}. 
This dataset contains MD trajectories of BRD4 in both the ligand-free form and 10 different ligand-bound forms (Tab.~\ref{tab:dg_comparison}), and the chemical structures of these ligands are shown in Fig.~\ref{fig:ligand_image}. 
Each system was simulated in three independent unbiased runs for 400~ns. 
We used the equilibrated trajectories after 50~ns and extracted only the protein atoms. 
Translation and rotation were removed from the protein trajectories by aligning the structure on the backbone atoms of 14 residues near the binding site to a reference protein structure \cite{yasuda2022}. 
Subsequently, the center of mass of the 14 residues was computed and included in the final dataset. 
The structure and dynamics of proteins in the dataset exhibit variations due to interactions with the binding ligand. 
In the previous work \cite{yasuda2022}, the short-term local dynamics of these 14 residues were suggested to show a pattern related to their ligand-binding free energy.

To examine the robustness of feature selection against changes in the candidate feature space, we further expanded the residue set from 14 to 19 residues by including additional residues surrounding the binding region (Fig.~\ref{fig:brd4_image}). 
The residues were selected such that any residue with a heavy atom within $6.5~\text{\AA}$ of the ligand in more than $50\%$ of the simulation frames was included.
This criterion ensures that the selected residues maintain frequent spatial proximity to the ligand during the trajectory. 
The same preprocessing procedure, including alignment and trajectory trimming, was applied to the 19-residue setting.

\subsection{Neural Network and Optimization}
The neural network consisted of two fully connected hidden layers with 2048 units each. 
The output layer was implemented without bias term. 
For each predefined mask size $n$, a separate network was trained. 
During training, randomly sampled masks were applied at each iteration, and optimization was performed for $5 \times 10^{5}$ iterations.

The Wasserstein distance was approximated using a gradient-penalized formulation with coefficient $\lambda = 10.0$. 
We used the Adam optimizer with parameters $\beta_1 = 0.0$, $\beta_2 = 0.9$, and $\epsilon = 10^{-3}$. 
The batch size during training was set to 256, and the learning rate was fixed at $\eta = 1 \times 10^{-4}$. 
During inference, a larger batch size of 2048 was used to reduce statistical fluctuations in the estimated quantities.

For mask optimization, 10 independent runs with different random seeds were performed with cooling rate $\gamma = 0.999$. 
Prior to simulated annealing, $10^{4}$ random masks were evaluated, and the mask yielding the lowest objective value was selected as the initial solution. 
Simulated annealing was conducted using the trained neural network to evaluate the objective function. 
The initial temperature was determined such that the acceptance probability of typical cost-increasing moves was set to 0.95.
To estimate the characteristic magnitude of the cost change $\Delta C$, 1000 random perturbations were evaluated, and the initial temperature was set as
\begin{equation}
    T_0 = - \frac{\mathbb{E}[\Delta C]}{\ln(0.95)} .
\end{equation}
In the simulated annealing procedure, an early-stopping criterion was introduced. The optimization was terminated when no improvement in the cost function was observed for 1000 consecutive steps ($K = 1000$), or when the maximum number of simulated annealing steps ($N_{\mathrm{SA}} = 50000$) was reached.

For low-dimensional embedding, the Wasserstein distance matrix was estimated by averaging the results of the last 1000 forward passes of the trained neural network with a batch size of 2048.
The embedding was optimized by minimizing Eq.~\ref{eq:embed} using simulated annealing followed by gradient-based refinement and setting the dimensionality of $\boldsymbol{p}$ to 3. 
To obtain a stable solution, 256 independent embedding trials were performed, and the embedding with the lowest objective value was selected as the final result.

\section{Results}
\subsection{Proof-of-concept automatic residue selection}
As a representative example, we present the results of auto-WHATMD when selecting four residues from the 14 binding-site residues.
In mask optimization, we computed the Wasserstein distance using a neural network trained with randomly sampled masks. 
While this approach does not rigorously compute the Wasserstein distance during the mask optimization process, we validated its effectiveness by comparing the Wasserstein distances obtained using random masks to those computed using the optimized masks. 
The random-mask approach yielded Wasserstein distances that correlated well with those from the optimized masks (Fig.~\ref{fig:random_fix_corr}), supporting
the validity of our approximation of the Wasserstein distance.

\begin{figure*}[h]
    %\vskip 0.2in
    \centering
    \includegraphics[width=150mm]{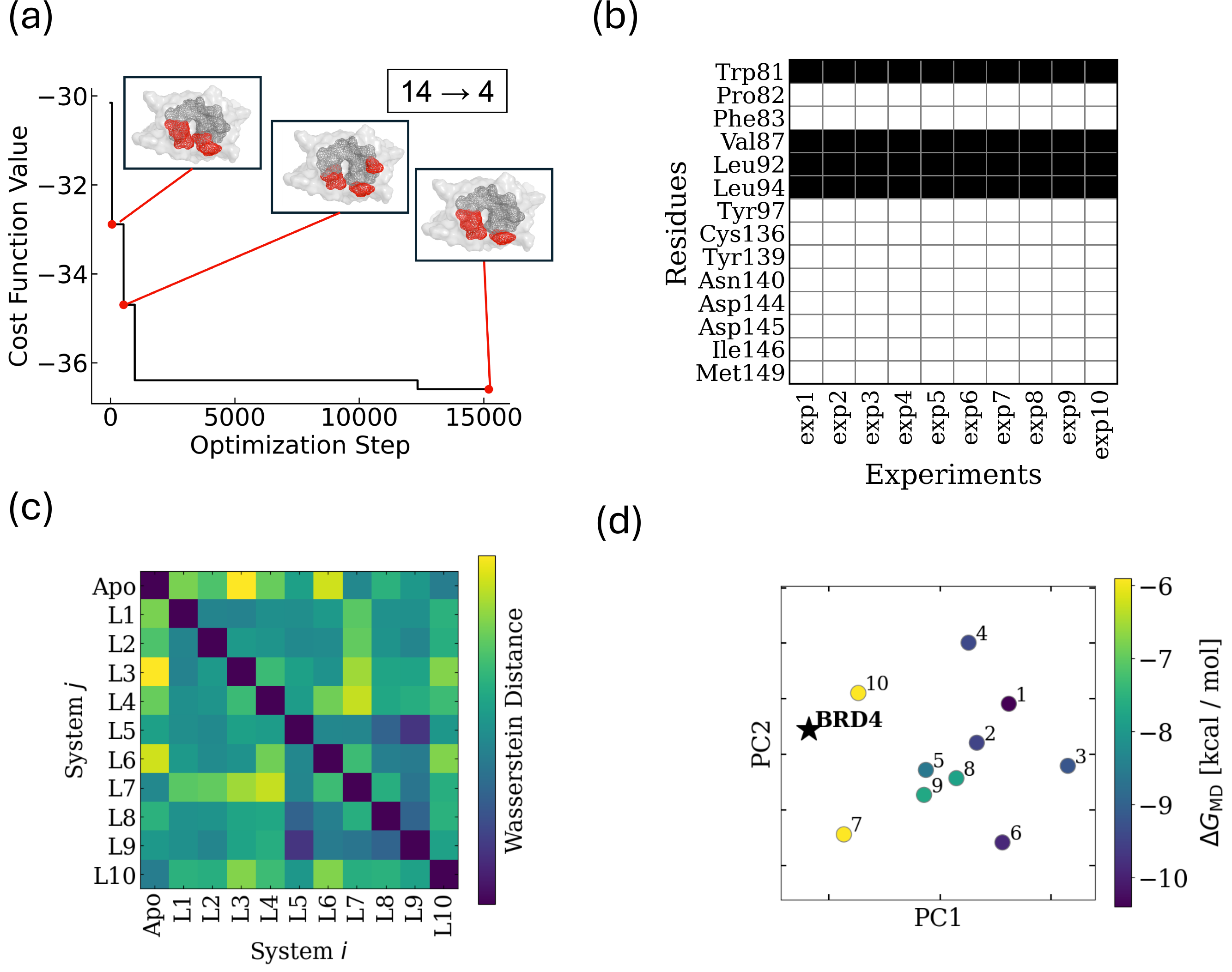} 
    \caption{
    Automatic selection from a small subset in BRD4 systems using auto-WHATMD.
    (a)~Representative process for optimizing residue selection, in which four residues were selected from a subset of 14 residues. 
    Selected residues are shown in red.  
    (b)~Residue selection results from ten independent optimization experiments. 
    Selected and non-selected residues are colored in black and white, respectively.
    (c)~Matrix of pairwise Wasserstein distances computed between all systems using the optimized binary mask. 
    Indices correspond to the ligand-free protein (apo) and the 10 ligand-bound systems (L1--L10). 
    (d) Principal component representation of the low-dimensional embedding derived from the Wasserstein distance matrix. 
    Each point represents a system labeled by ligand number. 
    The color indicates the computational ligand-binding free energy $\Delta G_{\mathrm{MD}}$ reported in Ref.~\cite{aldeghi2016}.
    }
    \label{fig:result_n4}
\end{figure*}

We first performed experiments to select $n$ residues, for $n = 4$ (Fig.~\ref{fig:result_n4}a). 
Ten independent optimization runs were conducted using simulated annealing.
Notably, Trp81, Val87, Leu92, and Leu94 were consistently selected in all trials (Fig.~\ref{fig:result_n4}b), demonstrating high stability and reproducibility of the optimization procedure.

The residues selected by our method are therefore consistent with existing knowledge regarding ligand-induced dynamical modulation.
Trp81 has previously been reported by NMR experiments to undergo ligand-induced dynamical changes~\cite{urick2016protein}, and its behavior has been suggested to correlate with binding affinity. In addition, Trp81, Val87, and Leu92 have been identified in prior computational and experimental studies as key residues influenced by ligand interactions. Leu92 and Leu94 are located near the hydrophobic binding pocket and are known to contribute to ligand stabilization through hydrophobic contacts~\cite{beier2024probing}. 
These residues were identified without incorporating any prior structural or biochemical information. 
%This result indicates that the proposed method can automatically extract amino acid residues whose dynamics are strongly affected by ligand binding. 
%Furthermore, the fact that identical residues were selected in all independent optimization runs further supports the robustness of the residue selection process.

Using the optimized mask vector (Trp81, Val87, Leu92, Leu94), we recalculated the pairwise Wasserstein distances between systems (Fig.~\ref{fig:result_n4}c). 
Relatively large distances were observed between the ligand-unbound system and ligand-bound systems, whereas distances among ligand-bound systems were comparatively small. 
This pattern suggests that the selected residues effectively capture the dynamical differences associated with ligand binding, while reflecting the similarity among bound states.

To further analyze the global structure of these dynamical differences, the Wasserstein distance matrix was embedded into a two-dimensional space for visualization (Fig.~\ref{fig:result_n4}d). 
The ligand-unbound system was clearly separated from the ligand-bound systems. 
Moreover, the first principal component (PC1) exhibited a monotonic relationship with the ligand binding free energies, indicating that the extracted dynamical feature is strongly associated with binding affinity.

\begin{figure}[t!]
    %\vskip 0.2in
    \centering
    \includegraphics[width=80mm]{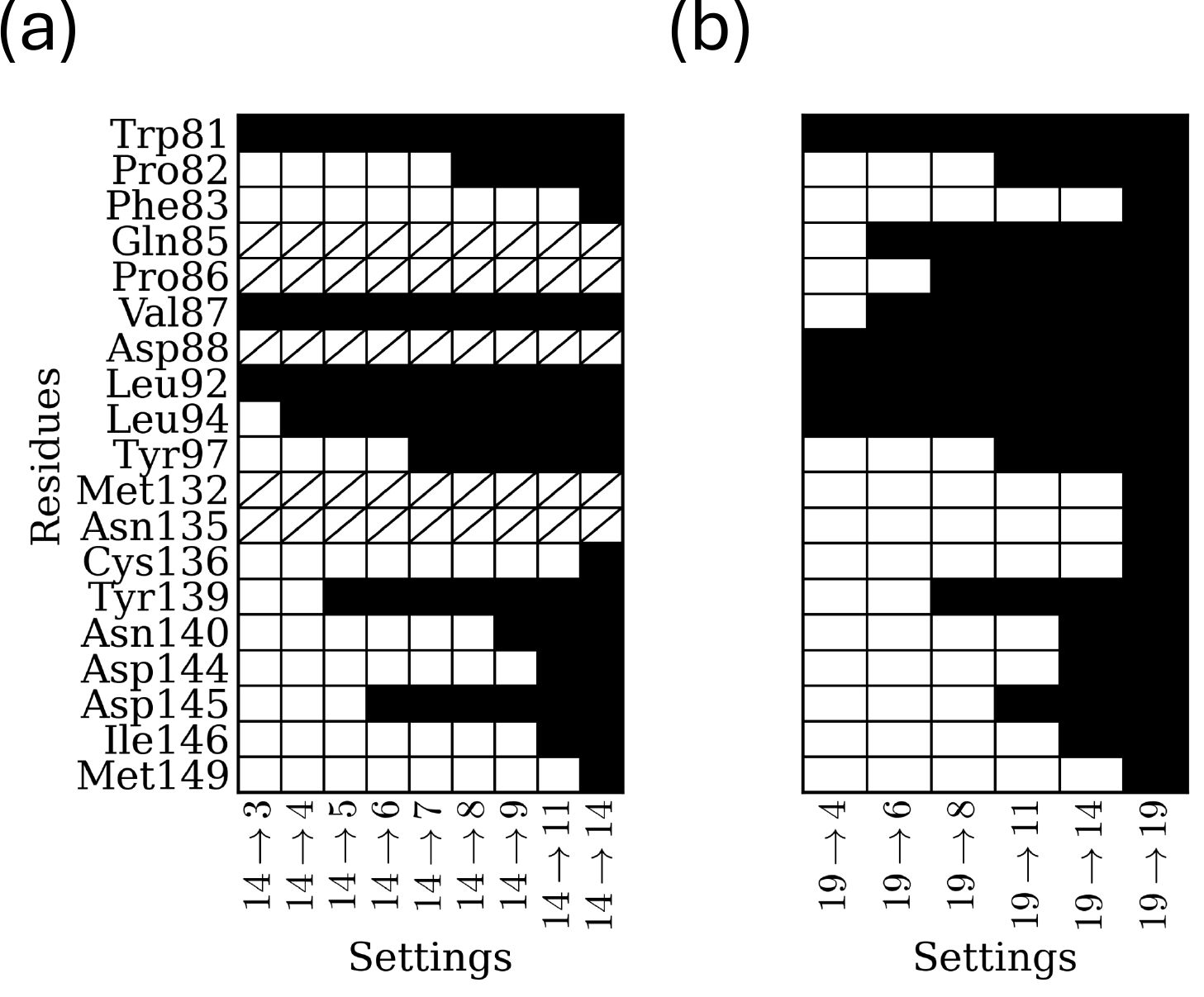}
    \caption{
    Consistency of selected residues via mask optimization for BRD4 systems. 
    (a) Results for the small subset of 14 residues, varying the number of selected residues from 3 to 14. Residues marked with a crossed line were not included in the subset. Selected and non-selected residues are colored in black and white, respectively. 
    (b) Results for the extended subset of 19 residues around the binding site, varying the number of selected residues from 3 to 19.
    }
    \label{fig:fig5_mask}
\end{figure}

\begin{figure}[t!]
    %\vskip 0.2in
    \centering
    \includegraphics[width=70mm]{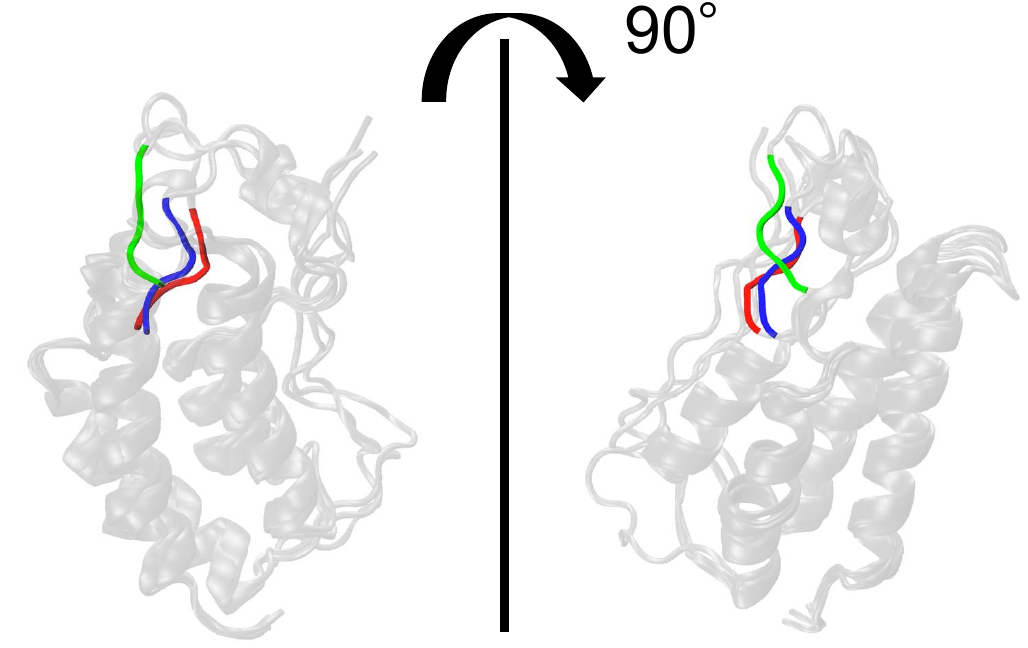}
    \caption{
    Representative ZA-loop conformations of the apo, L3-bound, and L10-bound systems. 
    The selected residues 85–88 are shown in red (the apo), blue (L3-bound), and green (L10-bound), while other residues are shown as transparent.
    }
    \label{fig:brd4_conf}
\end{figure}

\subsection{Robust residue selection across different subset sizes}

We further evaluated the behavior and robustness of the proposed method by (1) varying the number of residue selection in the 14-residues subset, (2) using an extended 19-residues subset. 
The 19 residues were defined as those exhibiting a contact frequency greater than 0.5 within 6.5~\AA{} of the ligand during the 10 ligand-bound simulations.

For the 14-residues subset, residues located in loop regions, such as Trp81, Val87, and Leu92, were selected when the number of selected residues was small (Fig.~\ref{fig:fig5_mask}a). 
As increasing the number of selected residues, residues around $\alpha$-helices were also selected.
In the 19-residue setting, the selected residue subset partially differed from that obtained using the 14-residue setting.
In addition to the previously identified residues Trp81, Val87, Leu92, and Leu94, the newly included residues located in the ZA loop region, namely Gln85, Val86, and Asp88, tended to be preferentially selected (Fig.~\ref{fig:fig5_mask}b).

Auto-WHATMD identified the ZA loop as a key region distinguishing the BRD4 systems, consistent with its known role in ligand recognition~\cite{wu2021molecular}. 
Previous studies have suggested that the ZA loop exhibits substantial conformational flexibility and can sample a range of conformations, particularly in the apo state, whereas ligand binding tends to restrict its conformational flexibility \cite{raich2021discovery, chen2022molecular}.
We compared representative conformations of the ZA-loop region (residues 85–88) (Fig.~\ref{fig:brd4_conf}). 
In the apo system, the ZA loop exhibits relatively flexible behavior and samples a broader region during the simulation. In contrast, in the L3-bound system, the ZA loop remains close to its initial position without large conformational changes.
The L10-bound system shows a different behavior from both the apo and L3-bound systems. 
While the ZA loop initially remains near the starting conformation similarly to the L3-bound system, it later adopts a conformation that is distinct from those observed in the apo and L3-bound trajectories, as illustrated in Fig.~\ref{fig:brd4_conf}.
These results indicate that auto-WHATMD detected differences of the ZA loop conformations.

\subsection{Relationship between a feature of selected residues and ligand-binding free energy}

As a measurement criterion of the embedding, we calculated Pearson's correlation coefficient between PC1 and the ligand binding energies. The ligand binding energies are an important indicator of the stability of protein--ligand complexes and range from \(-10.4\) to \(-5.9\) kcal/mol in this dataset, as reported in previous free energy calculations using MD simulations \cite{aldeghi2016} (Tab.~\ref{tab:dg_comparison}). 

\begin{figure}[t!]
    %\vskip 0.2in
    \centering
    \includegraphics[width=80mm]{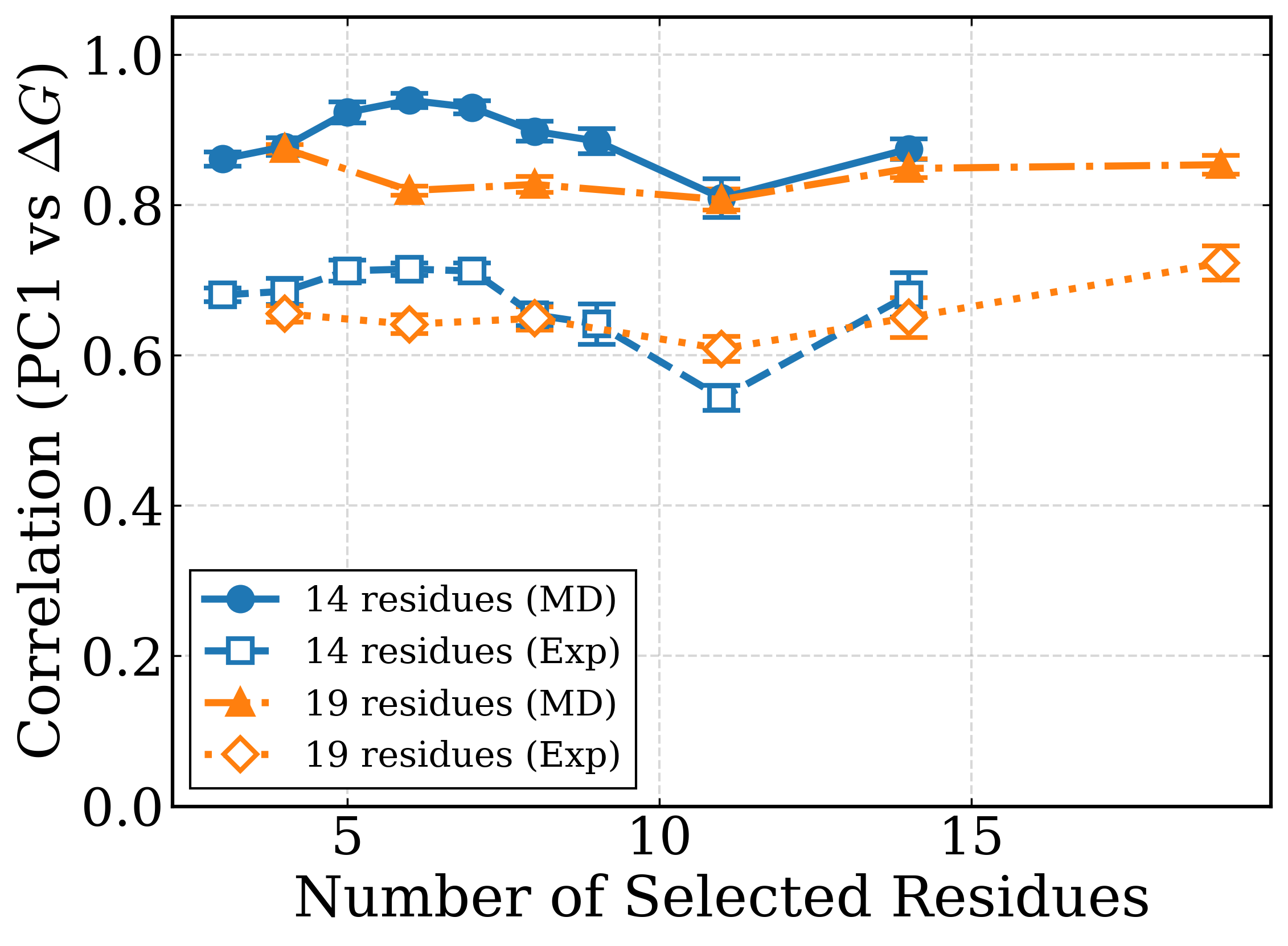}
    \caption{
    Correlation between ligand-binding free energy and the PC1 extracted from the Wasserstein distance matrix. Absolute Pearson correlation coefficients between PC1 and ligand-binding free energy ($\Delta G_{\mathrm{MD}}$ or $\Delta G_{\mathrm{EXP}}$) are shown as a function of the number of selected residues for the 14- and 19-residue subsets. Solid lines represent correlations with $\Delta G_{\mathrm{MD}}$, and dashed/dotted lines with $\Delta G_{\mathrm{EXP}}$.
    }
    \label{fig:fig5_corr}
\end{figure}

Using the 14-residues subset, the correlation to $G_{\rm{MD}}$ ranged from 0.77--0.94, depending on the number of selected residues (Fig.~\ref{fig:fig5_corr}; see also Figs. A2 and A3 for the corresponding two-dimensional embeddings in the PC1–PC2 space and PC1–$\Delta G$ plots, respectively).
The highest correlation with $G_{\rm{MD}}$ was observed at $n=6$. Similarly, the correlation with $G_{\rm{EXP}}$ ranged from 0.54--0.73 (Fig.~\ref{fig:fig5_corr}). 
We note that the lower performance with $G_{\rm{EXP}}$ may be attributed to the limited accuracy of the force field parameters and insufficient sampling in the MD simulations.  
Overall, these results suggest that the low-dimensional embedding obtained by the proposed method successfully extracts essential features associated with ligand binding.
%from high-dimensional and noise-containing MD simulation data.

The data representations between the 14- and 19-residues subsets differ slightly due to differences in the fitting references, but we then demonstrate that this fitting effect does not significantly effect PC1--$\Delta G$ correlation.  
For the 19-residues site, the Pearson correlation coefficient with $\Delta G_\mathrm{MD}$ ranged from 0.81--0.88, while the correlation with $\Delta G_{\mathrm{EXP}}$ ranged from 0.61--0.74 (Fig.~\ref{fig:fig5_corr}). 
These results indicate that high correlation values were preserved in the extended subset. 

\section{Limitations}
Although we have demonstrated that auto-WHATMD robustly selected key residues and that the feature based on the selected residues can dictate the ligand-binding properties, there are several limitations. 
First, we used the positions in the $xyz$ coordinate as inputs to represent the dynamics in three dimensional space, rather than commonly used features such as distance or angles as the input. 
We aligned all systems to a subset of residues in the reference structure to ensure translational and rotational invariance.
However, selecting reference residues is more challenging for flexible loops or when all residues of a protein are included as input, highlighting the need for improved residue representation. 
Second, our neural networks are trained separately for each pair of systems. 
Applying them to additional systems requires re-training the model for the new system pair.
Finally, our unsupervised learning framework does not ensure matching to the binding affinity. Combining the method with other features such as docking scores could increase the interpretability of the obtained features~\cite{igarashi2025protein}. 

\section{Conclusion}
We have presented auto-WHATMD, an algorithm for representing differences among multiple molecular dynamics trajectories. 
Applying it to BRD4 protein systems, we demonstrated that auto-WHATMD identified biologically important residues. 
The selection was robust regardless of the size of the candidate residue subset and the number of selected residues, ensuring consistent representation of system differences.
Moreover, the PC1 derived from auto-WHATMD correlated with ligand-binding free energies. 

We believe that auto-WHATMD effectively addresses the challenges associated with comparing ensembles. 
Protein dynamics are highly sensitive to various factors, such as ligand binding, temperature changes, and solvent types, which is a currently focused research area driven by extensive MD simulations. 
Auto-WHATMD could effectively characterize the differences between these systems, providing valuable biological insights into protein behavior.

%Overall, these results demonstrate that the proposed method maintains high predictive performance even when applied to higher-dimensional MD data. Moreover, depending on the scale of the search space, the method consistently extracts dynamical features ranging from local binding-related fluctuations to broader conformational differences that characterize system-dependent behavior within a unified analytical framework.

% \section*{Impact statement}
% This paper presents work whose goal is to advance the field of Machine Learning. There are many potential societal consequences of our work, none which we feel must be specifically highlighted here.


\bibliography{reference}
\bibliographystyle{icml2025}

%%%%%%%%%%%%%%%%%%%%%%%%%%%%%%%%%%%%%%%%%%%%%%%%%%%%%%%%%%%%%%%%%%%%%%%%%%%%%%%
%%%%%%%%%%%%%%%%%%%%%%%%%%%%%%%%%%%%%%%%%%%%%%%%%%%%%%%%%%%%%%%%%%%%%%%%%%%%%%%
% APPENDIX
%%%%%%%%%%%%%%%%%%%%%%%%%%%%%%%%%%%%%%%%%%%%%%%%%%%%%%%%%%%%%%%%%%%%%%%%%%%%%%%
%%%%%%%%%%%%%%%%%%%%%%%%%%%%%%%%%%%%%%%%%%%%%%%%%%%%%%%%%%%%%%%%%%%%%%%%%%%%%%%
\newpage
\appendix
\onecolumn
\renewcommand{\thetable}{A\arabic{table}}%
\setcounter{figure}{0}
\renewcommand{\thefigure}{A\arabic{figure}}

\section*{Appendix}
\begin{table}[h!]
    \centering
    \begin{tabular}{ccccc}
        \toprule
        \textbf{L} & $\Delta G_{\text{MD}}$ & $\Delta G_{\text{EXP}}$ & \textbf{PDB ID} & \textbf{Ref $\Delta G_{\text{EXP}}$} \\
        \midrule
        1 & $-10.4$ & $-9.8$ &  4OGI &  \cite{ciceri2014dual} \\
        2 & $-9.5$ & $-9.6$ & 3MXF &  \cite{filippakopoulos2010selective} \\
        3 & $-9.2$ & $-9.0$ & 4MR3 & \cite{picaud2013rvx} \\
        4 & $-9.4$ & $-8.9$ &  4OGJ &  \cite{ciceri2014dual} \\
        5 & $-8.6$ & $-8.8$ &  4J0R &  \cite{hewings2013optimization} \\
        6 & $-9.9$ & $-8.2$ &  3U5L &  \cite{filippakopoulos2012benzodiazepines} \\
        7 & $-5.9$ & $-7.8$ &  4MR4 &  \cite{picaud2013rvx} \\
        8 & $-7.8$ & $-7.4$ & 3U5J &  \cite{filippakopoulos2012benzodiazepines} \\
        9 & $-7.7$ & $-7.3$ & 3SVG &  \cite{picaud2013rvx} \\
        10 & $-5.9$ & $-6.3$ & 4HBV &  \cite{fish2012identification} \\
        \bottomrule
    \end{tabular}
    \caption{Calculated and experimental $\Delta G$ in kcal/mol values used in the dataset of BRD4 simulations. $\Delta G_{\rm{MD}}$ is obtained from \cite{aldeghi2016}.}
    \label{tab:dg_comparison}
\end{table} 

\begin{figure}[hb]
    \centering
    \includegraphics[width=60mm]{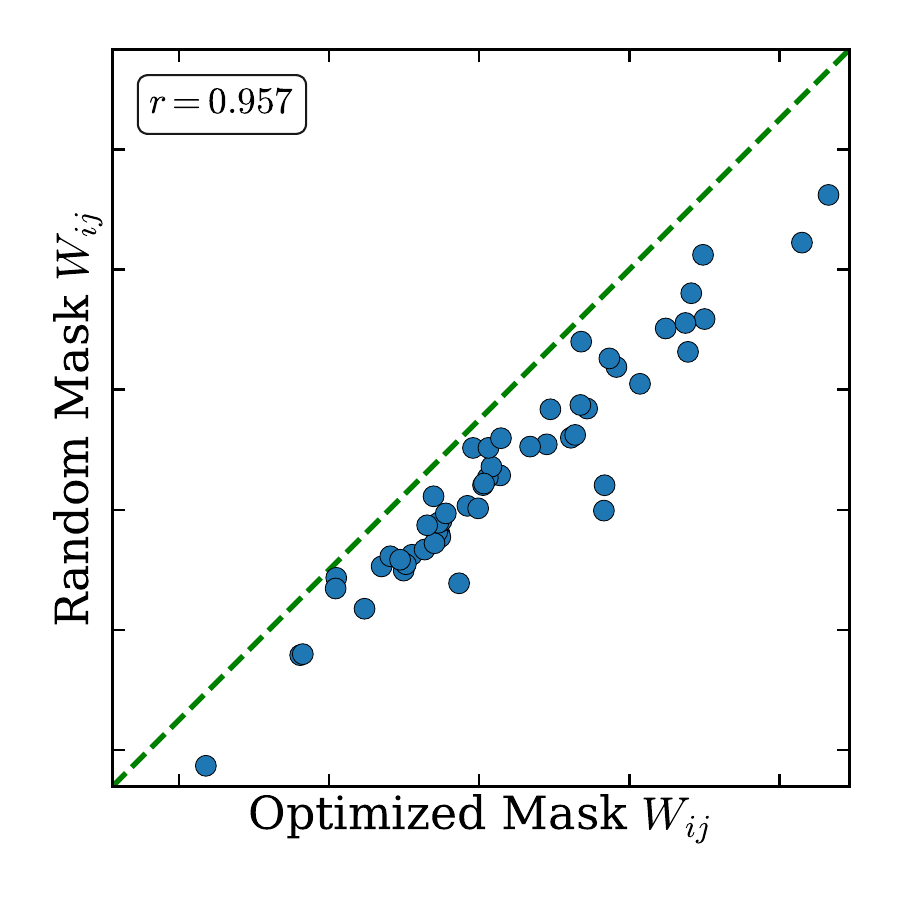}
    \caption{Comparison of the Wasserstein distance of all 55 pairs from the BRD4 systems, computed using neural networks trained with random masks (Random mask $W_{ij}$) and neural networks re-trained with the mask optimized for five-residues selection (Optimized mask $W_{ij}$). For optimized masks $W_{ij}$, the neural networks are initialized and re-trained for 500,000 steps. Green dashed line shows $y=x$.}
    \label{fig:random_fix_corr}
\end{figure}

\begin{figure}[tb]
    \centering
    \includegraphics[width=120mm]{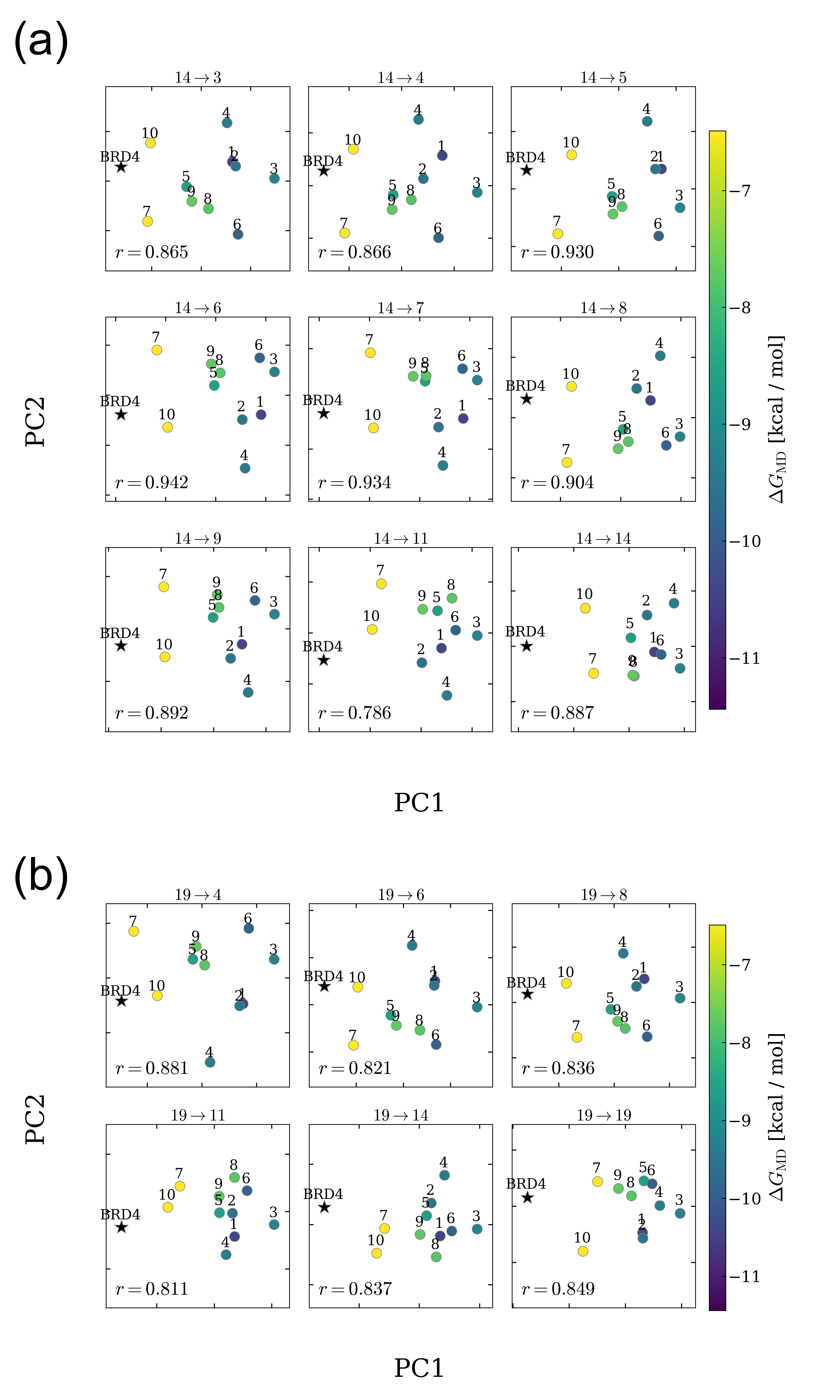}
    \caption{Two-dimensional embeddings of the BRD4 systems obtained from Wasserstein distance matrices computed from the selected residue subsets. The color of each point represents the binding free energy obtained from \cite{aldeghi2016}. (a) Results for the 14-residue candidate set, where the number of selected residues was varied from 3 to 14. (b) Results for the extended 19-residue candidate set around the binding site, where the number of selected residues was varied from 4 to 19.}
    \label{fig:brd4_pcs}
\end{figure}

\begin{figure}[tb]
    \centering
    \includegraphics[width=120mm]{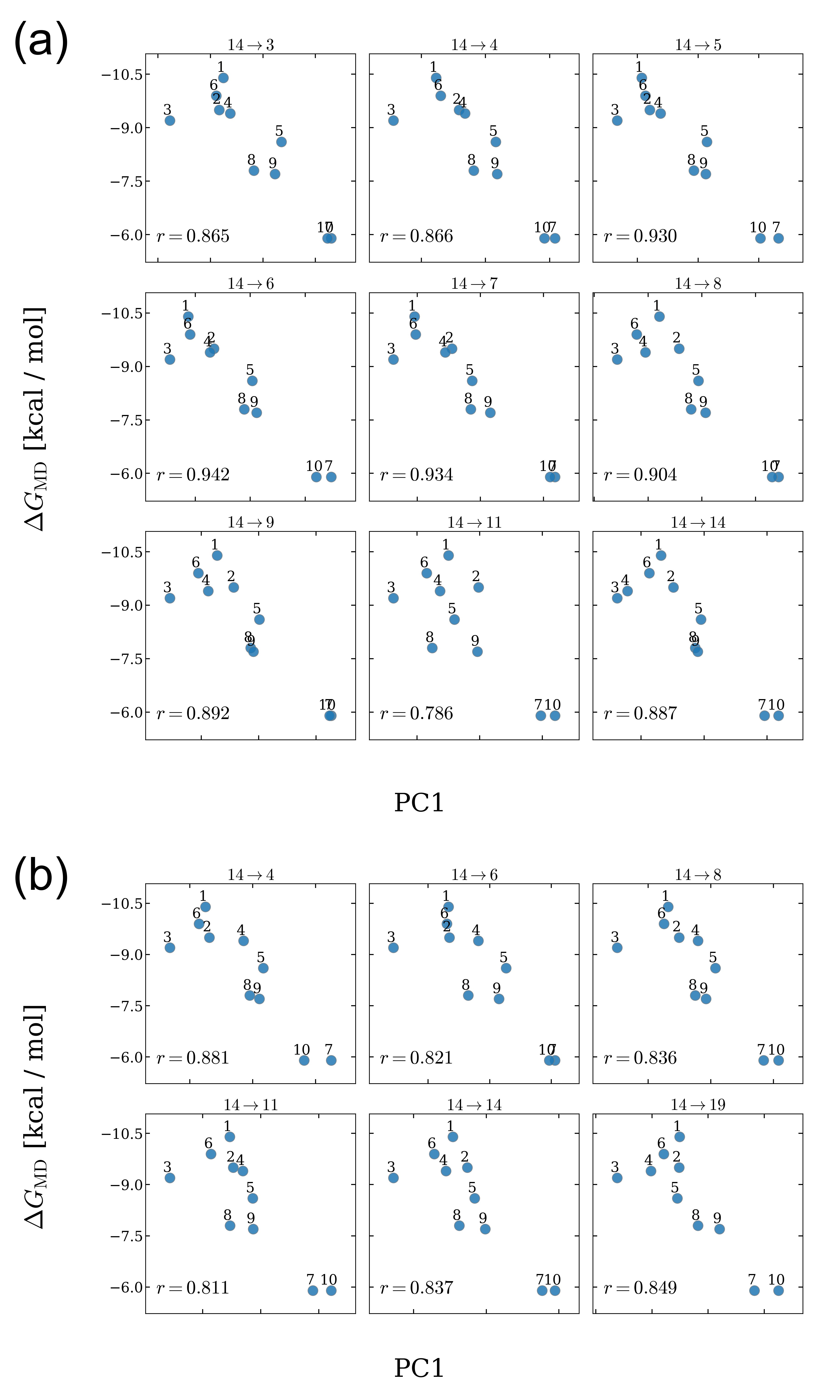}
    \caption{Plots of PC1 from auto-WHATMD with the binding free energy $\Delta G_{\mathrm{MD}}$ ~\cite{aldeghi2016} for the BRD4 systems. The numbers attached to the points indicate the ligand labels. (a) Results for the 14-residue candidate set, where the number of selected residues was varied from 3 to 14. (b) Results for the extended 19-residue candidate set around the binding site, where the number of selected residues was varied from 4 to 19.}
    \label{fig:brd4_pc1_vs_dgs}
\end{figure}

%You can have as much text here as you want. The main body must be at most $8$ %pages long.
%For the final version, one more page can be added.
%If you want, you can use an appendix like this one.  

%The $\mathtt{\backslash onecolumn}$ command above can be kept in place if you prefer a one-column appendix, or can be removed if you prefer a two-column appendix.  Apart from this possible change, the style (font size, spacing, margins, page numbering, etc.) should be kept the same as the main body.
%%%%%%%%%%%%%%%%%%%%%%%%%%%%%%%%%%%%%%%%%%%%%%%%%%%%%%%%%%%%%%%%%%%%%%%%%%%%%%%
%%%%%%%%%%%%%%%%%%%%%%%%%%%%%%%%%%%%%%%%%%%%%%%%%%%%%%%%%%%%%%%%%%%%%%%%%%%%%%%

\end{document}